\documentclass[fleqn,usenatbib,useAMS]{mnras}

\usepackage{graphicx}
\usepackage{epsfig}
\usepackage{rotating}
\usepackage{color}
\usepackage[normalem]{ulem}
\usepackage{multirow}
\usepackage{tabularx}
\usepackage{graphicx}
\usepackage{dcolumn}
\usepackage{hyperref}
\usepackage{float}
\usepackage{amsmath}
\usepackage{amssymb}
\usepackage{lipsum}
\usepackage{graphics}

\usepackage{lipsum}
\usepackage{mathtools}
\usepackage{cuted}
\usepackage{graphicx}	
\usepackage{amsmath}	
\usepackage{amssymb}	
\usepackage{multicol}        
\usepackage{bm}		
\usepackage{pdflscape}	

\newcolumntype{R}[1]{>{\raggedleft\arraybackslash}p{#1}}

\usepackage[T1]{fontenc}
\usepackage{ae,aecompl}

\usepackage{newtxtext,newtxmath}

\title[]{Modeling particle transport in astrophysical outflows and simulations of associated emissions}

\author[]{D. A. Papadopoulos,$^{1,3}$\thanks{E-mail: \href{mitsospapad@hotmail.com}{mitsospapad@hotmail.com} (DAP)}
O. T. Kosmas$^2$\thanks{E-mail: \href{odysseas.kosmas@manchester.ac.uk}{odysseas.kosmas@manchester.ac.uk} (OTK)}
and S. Ganatsios$^{3}$\thanks{E-mail: \href{ganatsios@teiwm.gr}{ganatsios@teiwm.gr} (SG)}
\\
$^1$Department of Electrical and Computer Engineering, University of Western Macedonia, Kozani, Greece\\
$^2$Modelling and Simulation Center, MACE, University of Manchester, Sackville Street, Manchester, UK\\
$^3$Theoretical Physics Section, University of Ioannina, GR-45110 Ioannina, Greece}
\date{Last updated 2020 June 10; in original form 2013 September 5}

\pubyear{2020}

\begin{document}
\label{firstpage}
\pagerange{\pageref{firstpage}--\pageref{lastpage}}
\maketitle

\begin{abstract}
In this work, after making an attempt to improve the formulation of the model on particle transport within astrophysical plasma 
outflows and constructing the appropriate algorithms, we test the reliability and effectiveness of our method through numerical 
simulations on well-studied Galactic microquasars as the SS 433 and the Cyg X-1 systems. Then, we concentrate on predictions of
the associated emissions, focusing on detectable high energy neutrinos and $\gamma$-rays originated from the extra-galactic M33 
X-7 system, which is a recently discovered X-ray binary located in the neighboring galaxy Messier 33 and has not yet been modeled 
in detail. The particle and radiation energy distributions, produced from magnetized astrophysical jets in the context of our 
method, are assumed to originate from decay and scattering processes taking place among the secondary particles created when hot 
(relativistic) protons of the jet scatter on thermal (cold) ones (p-p interaction mechanism inside the jet). These distributions 
are computed by solving the system of coupled integro-differential transport equations of multi-particle processes (reactions 
chain) following the inelastic proton-proton (p-p) collisions. For the detection of such high energy neutrinos as well as 
multi-wavelength (radio, X-ray and gamma-ray) emissions, extremely sensitive detection instruments are in operation or have been 
designed like the CTA, IceCube, ANTARES, KM3NeT, IceCube-Gen-2, and other space telescopes.
\end{abstract}





\section{Introduction}

During the last few decades, collimated astrophysical outflows have been observed to emerge from a wide variety of Galactic and 
extragalactic compact structures \citep*{Spencer_1984,Mirabel_Rodriguez_1999,Romero_Torres_et_al_2003,Reynoso08,Reynoso09,Romero_Review}. 
Among such objects, the  stellar scale class of microquasar (MQ) and X-Ray binary systems (XRBs) posses prominent positions 
\citep{Romero_Review,Smponias_tsk_2011,Smponias_tsk_2014,Romero-Vila-08,MJReid}. 
These two-body cosmic structures consist of a collapsed stellar remnant, a stellar mass black hole or a neutron star (compact 
object), and a companion (donor) main sequence star in coupled orbit around their center of mass. 

Due to the strong gravitational field pertaining around the compact object, mass from the companion star is accreted onto the 
equatorial region of the black hole forming an accretion disc. In the cases when the black hole is rotating rapidly and the 
accretion disc is geometrically rather thick and hot, ejection of two powerful oppositely directed mostly relativistic mass 
outflows (jets) occur perpendicular to the accretion disk \citep{Vieyro-Romero}. 
Such systems constitute excellent "laboratories" for the investigation of astrophysical outflows (jets). Currently, for the 
detection of multi-wavelength (radio, X-ray and gamma-ray) as well as high energy neutrino emissions, appreciably sensitive 
detection tools are in operation or have been designed like the IceCube, ANTARES, KM3NeT, CTA, IceCube-Gen2, etc.
\citep{MAGIC_Collaboration,Actis_Agnetta_Aharonian_2011,IceCube_2015,IceCube-PRL,IceCube-Science,KM3Net_2016,ANTARES-IceCube}. 
 
From a theoretical and phenomenological view point, due to the presence of rather strong magnetic fields, the MQ jets are 
treated as magneto-hydrodymamical flows emanating from the vicinity of the compact object (usually a stellar mass black hole)
\citep{Ody-Smpon-2015,Ody-Smpon-2017,Ody-Smpon-2018}, and this is also 
assumed to be the case in our present work. From the observed characteristics of MQs, researchers have concluded that they share 
a lot of similarities in their physical properties with the class of Active Galactic Nuclei (AGN), which are mostly located at 
the central region of galaxies. The latter, however, are enormously larger in scale compared to microquasars and, in addition, 
the evolution of AGNs is appreciably slower than that of MQs that makes the observation of many phenomena significantly 
difficult \citep{Remil,CharlesPA,Oros1,FabrikaS,CherepashchukAM}. 

In this article, we focus on the magnetized astrophysical outflows (jets) that are characterized by hadronic content (p, $\pi^\pm$, 
light nuclei, etc.) in their jets. We consider them as fluid flows emanating from a central source at the jet's origin 
\citep{Mirabel_Rodriguez_1999,Smponias_tsk_2011,Smponias_tsk_2014}. We, furthermore, assume that the compact object is a spinning 
black hole with mass up to few tens (30-50) of the Sun's mass \citep{Smponias_tsk_2014,Romero_Review,Reynoso08}.

Initially, we make an effort to improve the model employed to perform numerical simulations for particle and radiation emissions from 
hadronic MQ jets \citep{PapadD,PapOKos}. We adopt that the main mechanism producing high energy neutrinos and $\gamma$-rays, is the 
proton-proton (p-p) collision taking place within the hadronic jets, i.e. the inelastic scattering of hot (non-thermal) protons on 
thermal (cold) ones \citep{Reynoso08}. 

The scattering, diffusion, decay, etc., of the secondary particles ($\pi^{\pm}$, $K^{\pm}$, $\mu^\pm$, $e^\pm$, etc.) produced 
afterwards, from a mathematical modeling point of view, are governed by a system of coupled integro-differential transport equations. 
One of the purposes of this work is to attempt, for a first time, to formulate in a compact way this differential system of transport 
equations (satisfied by the primary protons and the secondary multi-particles, multi-species) and derive advantageous algorithms to 
perform the required simulations for the aforementioned emissivities \citep{Tsoulos_Ody_Stavrou}. 

Moreover, we are intended to extend the calculations of Ref. \citep{Smponias_tsk_2014,Ody-Smpon-2015,Ody-Smpon-2017,Ody-Smpon-2018} so 
as to include 
contributions to neutrino emissivity (intensity) originating from the secondary muons ($\mu^{\pm}$) produced from the decay of charged 
pions ($\pi^\pm$) which in our previous works had been ignored due to the long time consuming required for such simulations. Obviously, 
the mathematical problem of such a study becomes also more complicated (e.g. the number of coupled equations in the above mentioned 
differential system increases rapidly in the case of considering neutrinos coming out of kaon ($K^\pm$) decays \citep*{Lipari07}.

After fixing the model parameters and testing the derived algorithms on the reproducibility of some known properties of the well
studied SS 433 Galactic micro-quasar \citep{Ody-Smpon-2015,Ody-Smpon-2017,Ody-Smpon-2018}, we perform detailed simulations for the Galactic 
Cyg X-1 system as well as for the extragalactic M33 X-7 MQ \citep{PietschWC}. The latter system has not been studied up to now from a jet 
emission view point because this is a rather recently discovered X-ray binary system located in the neighboring galaxy of our Milky Way 
Galaxy, known as Messier 33 \citep{PietschWC}. 

The rest of the paper is organized as follows. In Sect. 2, we describe briefly the main characteristics of hadronic jets in Galactic 
micro-quasars 
(SS433 and Cyg X-1) as well as the extragalactic M33 X-7 system. Then (Sect. 3), we present the formulation of our improved method related 
to the differential system of transport  equations (integro-differential system of coupled equations) and derive the appropriate algorithms. 
In Sect. 4 we present and discuss our results referred to high energy
neutrino and $\gamma$-ray energy-spectra emitted from the extragalactic M33 X-7 MQ. 
Finally (Sect. 5), we summarize the main conclusions extracted from the present investigation.

\section{Brief description of the hadronic MQ systems }
\label{Brief description of MQ}


In this section we summarize briefly some basic properties of the hadronic MQ systems (like those mentioned before) and the
characteristics of hadronic models that describe reliably microquasar jet emissions 
\citep{Romero_Review,Smponias_tsk_2011,Smponias_tsk_2014,Vieyro-Romero}. 
The last few decades, from the investigation of a great number of MQs and X-ray emitting binary systems, stellar-mass black holes 
have been observed, with masses in the region of our interest determined mainly from the dynamics and structure properties of their 
companion stars \citep{Remil,CharlesPA,Oros1}. 

The Galactic binary system SS433, located 5.5 kpc from the Earth in Aquila constellation, displays two mildly relativistic jets (with 
bulk velocity $\upsilon_b \approx 0.26$c) that are oppositely directed and precess in cones \citep{JDRomney}. This system consists of a 
compact object (black hole) and an A-type companion (donor) star in coupled orbit with a period $P \sim 13.1$ days \citep{FabrikaS}. For 
the masses of the component stars of this system, in this work we adopted the observations of the INTEGRAL \citep{CherepashchukAM}, i.e. 
$M_{BH}=9M_{\odot}$ and $M_{don}=30M_{\odot}$ for the black hole and the companion star, respectively.

\begin{center}
\begin{table*}
\caption{Parameters of the model for M33 X-7, Cygnus X-1 and SS433}
\label{Table-1} 
\begin{tabular}{|l|c|l|l|l|}
\hline
&  &  &  &  \\
Parameter & Symbol  & SS433  & Cygnus X-1 & M33 X-7 \\  
\hline 
&  &  &  &  \\
Black Hole mass                & $M_{BH}$    & $9.0M_{\odot}$ & $14.8M_{\odot}$ & $15.65M_{\odot}$ \\
Distance from Earth            & $d$         & 5.5 kpc    & 1.86 kpc   & 840-960 kpc \\
Donor Star mass                & $M_{don}$   & $30 M_{\odot}$    & $19.2 M_{\odot}$   & $ 70 M_{\odot}$ \\
Donor star type                & -           & A-type     &O-type  & O-type \\
Orbital Period                 & $P$         & 13.1 days  & 5.6 days   & 3.45 days  \\
Jet's kinetic power            & $L_k$       & $10^{39}erg$ $s^{-1}$ & $10^{38}erg$ $s^{-1}$ & $10^{38}erg$ $s^{-1}$ \\
Jet's launching point          & $z_0$       & $1.3\times 10^9$ cm & $10^8$ cm & $1.9\times 10^8$ cm \\
Bulk velocity of jet particles & $\upsilon_b$& 0.26c & 0.6c  & 0.8c  \\
Jet's bulk Lorentz factor      & $\Gamma_b$  & 1.04  & 1.25  & 1.66  \\
Jet's half-opening angle       & $\xi$       & $0.6^{\circ}$  & $1.5^{\circ}$ & $7^{\circ}$ \\
Jet's viewing angle            & $\theta$    & $78.05^{\circ}$ & $27.1^{\circ}$ & $74.6^{\circ}$ \\ 
\hline 
\end{tabular}
\end{table*}
\end{center}

The second Galactic system addressed in this work, Cygnus X-1, is an X-ray source in the constellation Cygnus which is located 
1.86 kpc from the Earth \citep{MJReid}. The $19.2 M_{\odot}$ O-type companion star \citep{Orosz2011} is in coupled orbit with the 
$14.8 M_{\odot}$ 
compact object (stellar mass black hole) \citep{Orosz2011}, with orbital period $P \sim 5.6$ days \citep{Gies2008}. 

As mentioned before, one of our main goals in this work is to calculate high energy $\gamma$-ray and neutrino emissions from 
the recently discovered binary system M33 X-7 \citep{Long} which is located in the nearby galaxy Messier 33, the only known up to 
now black hole that is in an eclipsing binary system, with orbital period $P \sim 3.45$ days \citep{PietschWC}. The distance from 
Earth of this system is between $\sim 840$ kpc \citep{Oros2} and $\sim 960$ kpc \citep{Bonanos}. In addition, the black hole mass 
is $M_{BH}=(15.65\pm 1.45) M_{\odot}$ which orbits its $M_{don}=(70.0 \pm 6.9) M_{\odot}$ O-type companion star. 

In Table \ref{Table-1}, we tabulate in addition some other important parameters employed in this work for the systems under 
investigation, SS 433, Cygnus X-1 and M33 X-7.

\subsection{ Dynamics and energetic evolution of hadronic MQ jets }

In the model considered in this work for the description of MQ jets, an accretion disk in the equatorial region of the compact 
object is present, and a fraction of the accreted material is expelled in two oppositely directed jets 
\citep{Falcke,Smponias_tsk_2011,Smponias_tsk_2014}. We assume an approximately conical mass outflow (jet) with a half-opening angle 
$\xi$ (for the M33 X-7 MQ, we adopt the value $\xi=7^{\circ}$) and a radius given by $r(z)=z\tan\xi$ (the coordinate z-axis coincides
with the cone axis, jet's direction). The injection point (plane) of the jet is at a distance $z_0$ from the compact object. 
When $z=z_0$ the radius of the jet is given by $r_0=z_0\tan\xi$ (see Table I). 

The initial jet radius is $r_0=r(z_0) \approx 5R_{sch}$, where $R_{sch}=2GM_{BH}/c^2$ ($R_{sch}$ is the known Schwarzschild radius, 
the critical black hole radius for which the escape speed of particles becomes equal to the speed of light $c$), with $G$ 
denoting the Newton's gravitational constant. Then, we find, for example, that the injection point is at $z_0=r_0/\tan\xi\simeq 1.9 \times 10^8$ 
cm, for the M33 X-7. For the sake of comparison, we mention that for Cygnus X-1, the injection point is at $z_0\simeq 10^8$ cm, 
while for SS 433 it is $z_0\simeq 1.3\times 10^9$ cm. 

It should be also noted that, for all systems studied in this work, the extension of the jet is determined through a maximum value 
of z, denoted by $z_{max}$, which is assumed to be equal to $z_{max}=5 z_0$ (this point may be assumed to be at the boundaries of 
the jet with the ambient region).

In discussing the energetic evolution and dynamics of MQ jets, the kinetic energy density of the jet, $\rho_k(z)$, is related 
to its kinetic luminosity, $L_k$, through the expression \citep{Romero_Review}
\begin{equation}
\rho_k(z)=\frac{L_k}{\pi \upsilon_b {[r(z)]}^2} \, ,
\label{Kin-Lumin}
\end{equation}
where $\upsilon_b$ is the bulk velocity of the jet particles (mostly protons). Within the context of the jet-accretion coupling 
hypothesis, only around $10 \%$ of the Eddington luminosity goes into the jet \citep{Kording}. Here we adopt $L_k=10^{38}erg \ s^{-1}$,
for a black hole of mass $\sim 15.7 M_{\odot}$, i.e. for the M33 X-7 system. 

Furthermore, assuming equipartition between the magnetic energy and the kinetic energy in the jet, it implies that $\rho_{mag}=\rho_k$ 
\citep{Bosch}, and hence the magnetic field that colimates the astrophysical jet-plasma is given by \citep{Reynoso09}
\begin{equation}
B(z)\equiv \sqrt{8\pi\rho_{mag}(z)} = \sqrt{8\pi\rho_k(z)} .
\label{MagnField}
\end{equation}
In general, for the kinetic power in the jet, many authors consider that a fraction is carried by primary protons and the rest by electrons, 
i.e., $L=L_p+L_e$. Then, the relation between the proton and electron power is determined through a parameter $\alpha$ in such a way 
that $L_p=\alpha L_e$ \citep{Reynoso08,Reynoso09,Vieyro-Romero}. 

In this article, however, we adopt the case of $\alpha=100$ which means that we consider proton-dominated jet. Other quantities needed 
for the purposes of this paper are listed in Table I. 

\subsection{p-p Collision Mechanism inside MQ jets}

The collision of relativistic protons with the cold ones inside the jet (p-p collision mechanism), produces high energy charged particles 
(pions $\pi^\pm$, kaons $K^\pm$, muons $\mu^\pm$, etc.) and neutral particles ($\pi^0$, $K^0$, ${\tilde K}^0$, $\eta$ particles, etc.). 
Important primary reactions of this type are 
\begin{equation}
\begin{split}
p+p & \rightarrow p+p +          a\pi^0 + b(\pi^+ + \pi^-) \, , \nonumber \\ 
p+p & \rightarrow p+n +  \pi^+ + a\pi^0 + b(\pi^+ + \pi^-) \, , \\
p+p & \rightarrow n+n + 2\pi^+ + a\pi^0 + b(\pi^+ + \pi^-) \, , \nonumber   
\end{split}
\end{equation}
where $a$ and $b$ denote the pion multiplicities \citep{Romero_Review}, and similarly for kaon production \citep{Lipari07}.  

Charged pions $\pi^\pm$ (and kaons $K^\pm$), afterwards, decay to charged leptons (muons $\mu^\pm$, and electrons or positrons, 
$e^\pm$) as well as neutrinos and anti-neutrinos as
\begin{equation}
\begin{split}
&\pi^+ \rightarrow \mu^+ +\nu_{\mu}, \quad \quad \pi^+ \rightarrow e^+ + \nu_e \\
&\pi^- \rightarrow \mu^- + \bar{\nu_{\mu}}, \quad \quad \pi^- \rightarrow e^- + \bar{\nu_e} \, .
\end{split}
\label{pi-decay}
\end{equation}
Furthermore, muons also decay giving neutrinos and electrons (or positrons) as
\begin{equation}
\mu^+\rightarrow e^+ +\nu_e +\bar{\nu_{\mu}} , \quad \quad \mu^-\rightarrow e^- +\bar{\nu_e} + \nu_{\mu} \, .
\label{mu-decay}
\end{equation}

We mention that, neutral pions ($\pi^0$), $\eta$-particles, etc. decay producing $\gamma$-rays according to the reactions
\begin{equation}
\pi^0 \rightarrow \gamma + \gamma \, , \quad \quad \pi^0 \rightarrow \gamma + e^- + e^+ 
\label{pi-zero-decay}
\end{equation}
\begin{equation}
\eta \rightarrow \gamma + \gamma \, , \quad \quad \eta \rightarrow \pi^0 + e^- + e^+ \, .
\end{equation}
In our present study, we consider neutrinos generated from both types of charged-pion decays (for neutrinos coming from kaon 
decays the reader is referred, e.g. to Ref. \citep{Lipari07} and also from both types of charged muon decays \citep{PapadD,PapOKos}.
Moreover, we consider $\gamma$-rays produced from the $\pi^0$-decays of Eq. (\ref{pi-zero-decay}). Thus, high-energy neutrinos 
are inevitably accompanied by pionic gamma-rays, a phenomenon well known as multi-messenger emission from microquasar jets.

\subsection{ Accelerating and cooling rates of the jet processes }

In general, the primary charged particles (p and $e^-$) gain energy during moving within the magnetic field $B$ (Fermi
acceleration). The acceleration rate of the (initially cold) protons to an energy $E=E_p$, known as shock acceleration, is 
defined as $t_{acc}^{-1}=E^{-1}dE/dt$ and is given by the relation \cite{}
\begin{equation}
t_{acc}^{-1} \equiv \frac{1}{E} \frac{dE}{dt} \approx \eta \frac{c e B}{E_p}
\end{equation}
($\eta=0.1$ is the acceleration efficiency which means that only 10\% of the cold/thermal protons may acquire relativistic
energies). This is equivalent to existence of an efficient accelerator at the base of the jets where shocks are rather
relativistic \citep{Begelman_et_al_1980}.

In the p-p collision mechanism, the cross section and the rate of inelastic p-p scattering (relativistic protons scatter 
on cold ones), $\sigma_{pp}^{inel}$ and $t_{pp}^{-1}$ respectively, play crucial role (the in-elasticity coefficient is 
$K_{pp}\approx 1/2$). The corresponding cross section, $\sigma_{pp}^{inel}$, as a function of the fast proton energy $E_p$, 
reads 
\citep{Kelner1,Kelner2}
\begin{eqnarray}
\sigma_{pp}^{inel}(E_p)=(0.25L^2+1.88L+34.3)&\left[1-\left(\frac{E_{th}}{E_p}\right)^4\right]^2\nonumber\\
&\times 10^{-27} cm^2
\end{eqnarray}
where $L=ln(\frac{E_p}{1000 \quad GeV})$, and $E_{th}$ denotes the minimum energy of (thermal) protons, $E_p^{min}$, which 
is equal to $E_p^{min}\equiv E_{th}=1.22$ GeV, see e.g. Appendix of Ref. \citep{Smponias_tsk_2014}. 

The corresponding rate, $t_{pp}^{-1}$, of the inelastic p-p scattering is given in Table \ref{Table-2} in terms of the 
cross section $\sigma_{pp}^{inel}$ and the number density of cold particles (jet protons), $n_p(z)$, at a distance $z$ 
from the black hole which may be written as \citep{}
\begin{equation}
n_p(z)=\frac{(1-q_{rel})}{ \Gamma_b m_p c^2 } \rho_k (z) \, ,
\end{equation}
($q_{rel}$=0.1 denotes the portion of the relativistic protons). 

In Table \ref{Table-2}, in addition to the p-p collision rate and accelerating rates of protons, we tabulate the most 
significant cooling rates (synchrotron,adiabatic, etc.) for protons, pions and muons as well.

\begin{center}
\begin{table*} 
\caption{Accelerating, cooling and decay rates for particles moving inside hadronic MQ jets} 
\label{Table-2}
\centering
\begin{tabular}{|l|c|l|}
\hline
&  &  \\
Rate Parameter & Rate Symbol & Basic rate definition \\  
\hline
&  &  \\
Proton accelerating rates    & $t_{acc}^{-1}$  & $\eta \frac{ceB}{E_p}$ \\ 
Proton-proton collision rate & $t_{pp}^{-1}$   & $n_p(z) \sigma_{pp}^{inel} (E_p) K_{pp}$ \\
Synchrotron radiation rate  & $t_{syn}^{-1}$ & $\frac{4}{3}{\left(\frac{m_e}{m}\right)}^3 \frac{\gamma \sigma_T B^2}{m_e c 8\pi}$ \\
Adiabatic expansion rate & $t_{ad}^{-1}$ & $\frac{2}{3} \frac{\upsilon_b}{z}$ \\
Pion-proton collision & $t_{\pi p}^{-1}$ & $0.5 \ c \ n_p(z) \sigma_{\pi p}^{inel}(E_p)$   \\
Proton escape rate & $t_{esc}^{-1}$  & $ c/(z_{max}-z)$  \\
Pion decay rate & $t_{\pi}^{-1}$  & $t_{esc,\pi}^{-1} + t_{dec,\pi}^{-1}$ \\
Muon decay rate & $t_{\mu}^{-1}$  & $t_{esc,\mu}^{-1} + t_{dec,\mu}^{-1}$  \\
\hline 
\end{tabular}
\end{table*}
\end{center}

In more detail, the well known synchrotron radiation is emitted from charged particles of energy $E=\gamma m c^2$,
with $\gamma$ being the particle's Lorentz factor and $m$ its mass, moving inside the magnetized plasma. Thus, the 
rate of the synchrotron radiation, $t_{syn}^{-1}$, is very important in MQ systems with strong magnetic fields (see 
Table 2).

Furthermore, due to the adiabatic expansion of the jet plasma, all co-moving particles inside it loose energy with 
a rate $t_{ad}^{-1}$ \citep{Bosch}. This rate, known as adiabatic cooling rate, depends on the particle velocities 
$\upsilon_b\tan\xi$. 

In addition, $t_{\pi p}^{-1}$ gives the rate of pion-proton ($\pi-p$) inelastic scattering inside the jet. The corresponding 
$\sigma_{\pi p}^{inel}$ cross section is related to the inelastic p-p scattering cross section with the relationship 
$\sigma_{\pi p}(E)=(2/3)\sigma_{pp}^{inel}(E)$, where the factor $2/3$ comes out of the fact that protons are made of three 
valence quarks, while the pions of only two quarks \citep{Gaisser}.

It is worth mentioning that some processes taking place inside the jet, lead to particles' knock out, either by
escaping from the jet (and/or from the energy range of interest) or by their decay processes
\citep{Ginzburg}. Thus, the corresponding rates are: (i) the escaping rate, $t^{-1}_{esc}$, given by
\begin{equation}
t_{esc}^{-1} \approx \frac{c}{z_{max}-z} \, ,
\label{t_esc}
\end{equation}
(${z_{max}-z_0}$ represents the length of the acceleration zone), and (ii) the decay rate of the particle in question,
$t^{-1}_{dec}$, which is known from direct measurements of particle's life time. 

The rates of the knock out (or catastrophic) processes inside the jet for pions, $t_\pi^{-1}$, and muons, $t_\mu^{-1}$, 
include contributions from the decay process and escaping process as
\begin{equation}
t_{\pi}^{-1}(E,z)=t_{esc,\pi}^{-1}(z)+t_{dec,\pi}^{-1}(E) \, ,
t_{\mu}^{-1}(E,z)=t_{esc,\mu}^{-1}(z)+t_{dec,\mu}^{-1}(E) \, ,
\label{pion-muon-rate}
\end{equation}
where 
$ t_{dec,\pi}^{-1}= {[(2.6\times 10^{-8}) \gamma_{\pi}]}^{-1} \ s^{-1}$, 
for pions, and 
$t_{dec,\mu}^{-1}={[(2.2\times 10^{-6}) \gamma_{\mu}]}^{-1} \ s^{-1} $, 
for muons. We note that Eq. (\ref{pion-muon-rate}) holds also for protons due to the fact that $t^{-1}_{dec,p}=0$ (or 
equivalently the proton life time is infinite). 

The rates discussed above enter the set of basic coupled transport equations (system of kinematic equations) that describe
the particle distributions as we will discuss in Section 3. 

\subsubsection{ Energy variation of the various rates of jet particles }

\begin{figure}
\includegraphics[width=0.5\linewidth]{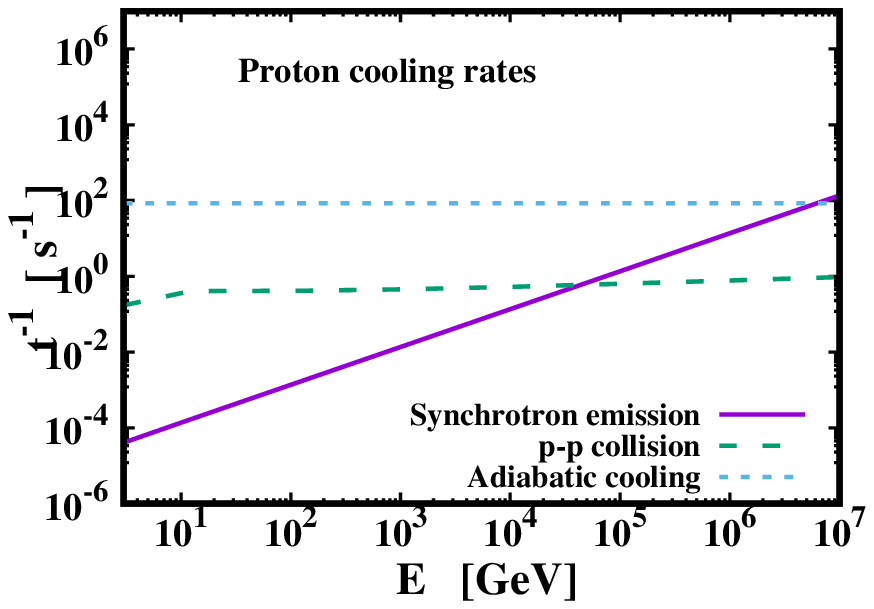} \\
\includegraphics[width=0.5\linewidth]{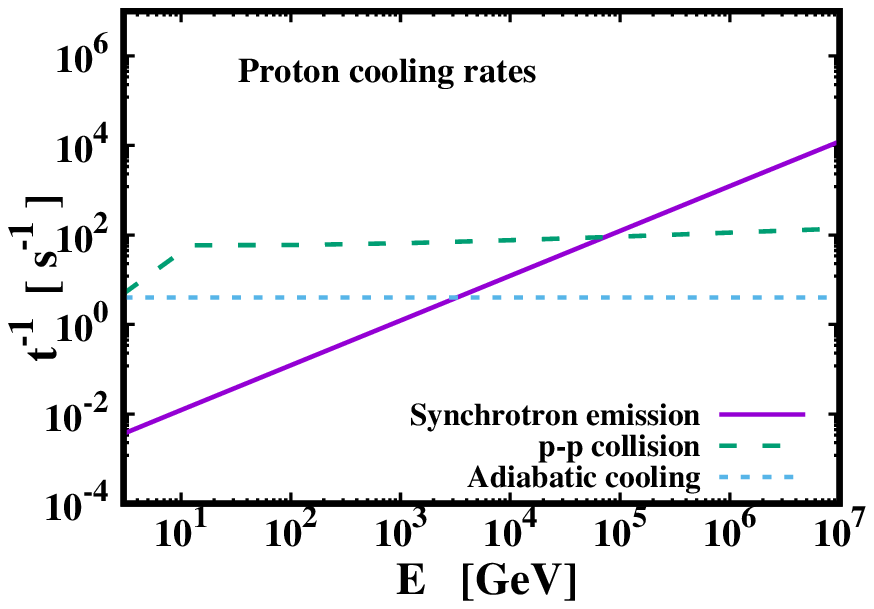}
\vspace*{0.2 cm}
\caption{ Cooling rates for protons in the jets of M33 X-7 (top) and SS433 (bottom) at the base of the jets $z_0$. The plots 
of cooling rates show the synchrotron emission (solid lines), the adiabatic cooling (dotted lines), and the 
inelastic p-p collision rate (dashed lines). }
\label{Fig-1}
\end{figure}

\begin{figure}
\includegraphics[width=0.5\linewidth]{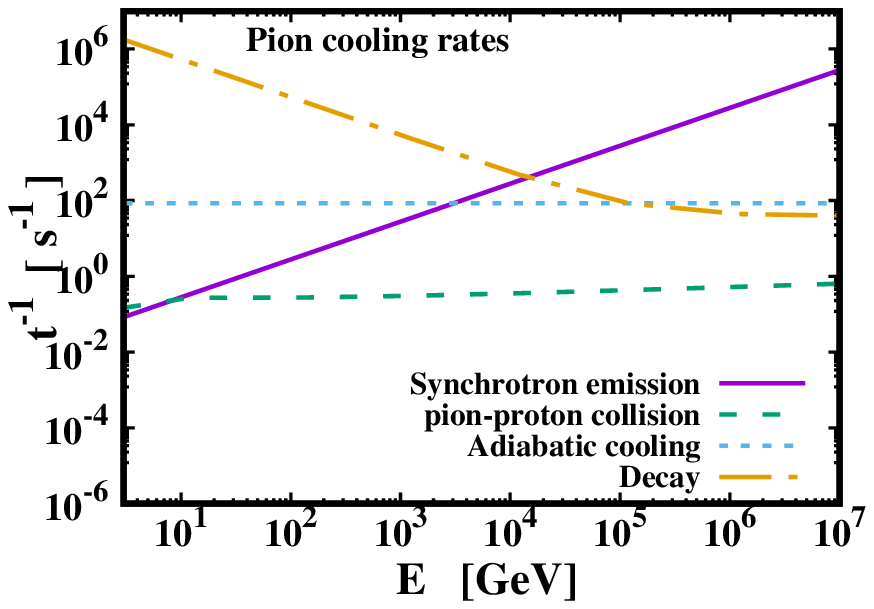} \\
\includegraphics[width=0.5\linewidth]{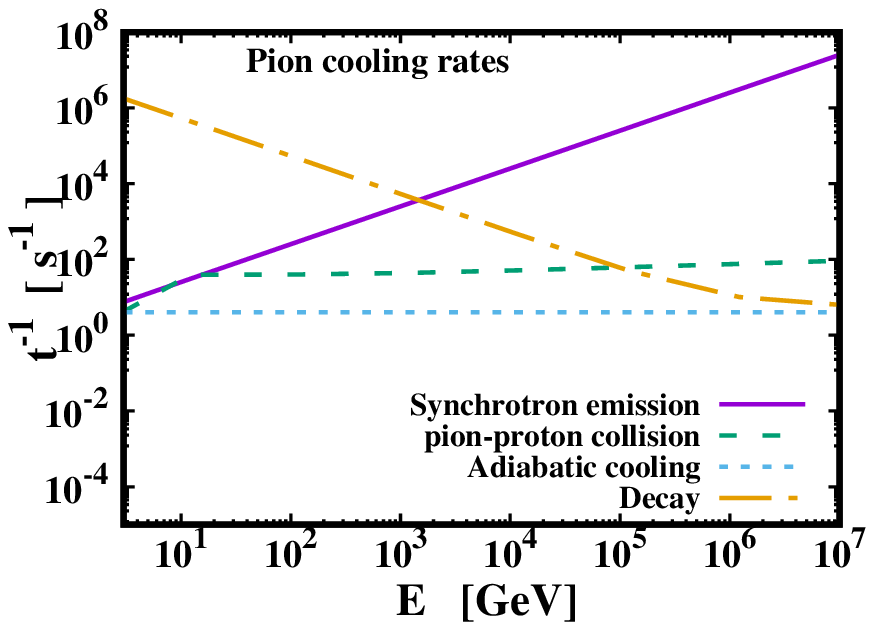}
\vspace*{0.2 cm}
\caption{Cooling rates for pions in the jets of M33 X-7 (top) and SS433 (bottom) at the base of the jets}
\label{Fig-2}
\end{figure}

\begin{figure}
\includegraphics[width=0.5\linewidth]{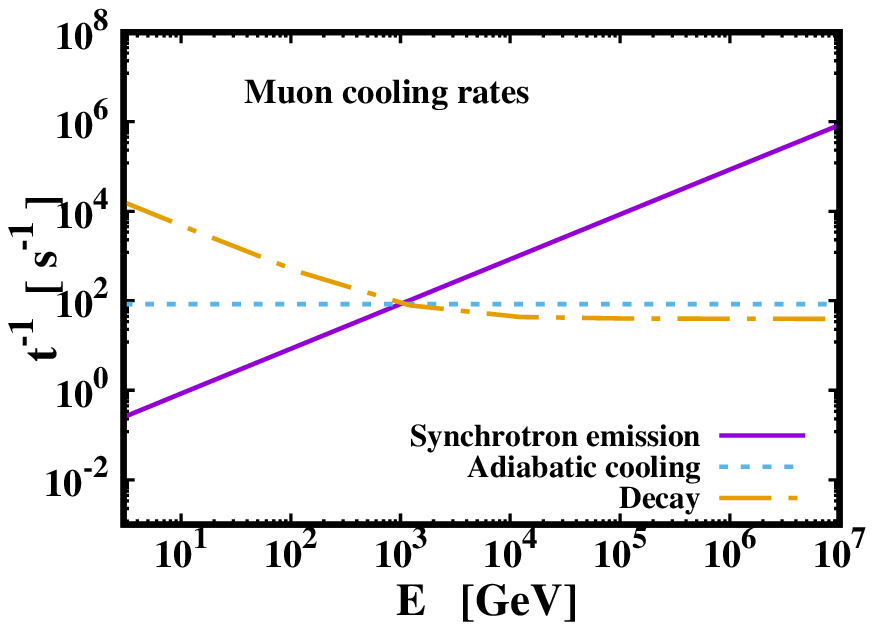} \\
\includegraphics[width=0.5\linewidth]{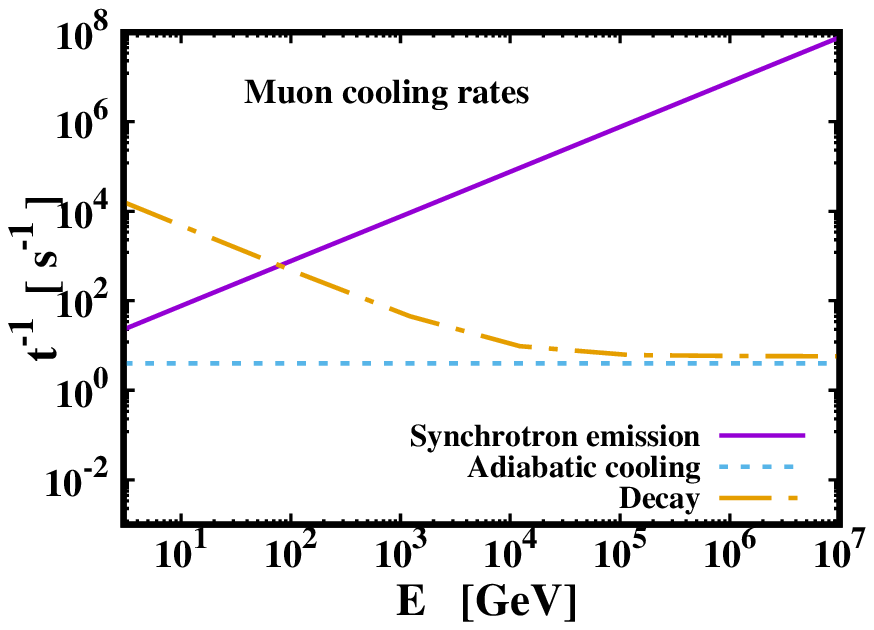}
\vspace*{0.2 cm}
\caption{Cooling rates for muons in the jets of M33 X-7 (top) and SS433 (bottom) at the base of the jets}
\label{Fig-3}
\end{figure}

In Figs. \ref{Fig-1}, \ref{Fig-2} and \ref{Fig-3}, the variation of the aforementioned main rates, versus the particle energy 
E, for protons, pions and muons, respectively, and for the SS 433 and M33 X-7 MQ systems, are illustrated. In other words, in 
these figures the region of dominance of the accelerating, cooling, etc. rates throughout the energy range of interest, 
$E_{th}\equiv E^{min}\le E\le E^{max}$, are demonstrated.

More specifically, Fig. \ref{Fig-1} indicate the cooling rates at the base of the jet for protons, Fig. \ref{Fig-2} those 
for pions and Fig. 3 the corresponding ones for muons, in the extragalactic binary system M33 X-7 (top of these figures). 
For comparison, the corresponding results for the Galactic binary system SS 433 (bottom of these figures) are also shown. 

The plots in Fig. 2 for pions, show respectively the variation versus the energy $E$ of the rate for synchrotron emission 
(solid lines), for the pion-proton inelastic collision (dashed lines), for the adiabatic cooling (dotted lines), and for 
the decay of pions (dot-dashed lines). They refer to the M33 X-7 (top) and the SS 433 (bottom) micro-quasar systems.
Also, Fig. 3 illustrates the cooling rates of muons for the synchrotron emission (solid lines), the adiabatic expansion
(dotted lines), and for the decay of muons (dot-dashed lines). They are similar to those for pions (Fig. 2) except
the corresponding muon-proton and muon-pion scattering channels which are ignored in this work.

From Figs. \ref{Fig-2} and \ref{Fig-3} it becomes obvious that, the particle synchrotron losses dominate the high energy 
region. In the case of protons (Fig. \ref{Fig-1}), however, due to the large proton mass, synchrotron losses are not 
dominant up to very high energies ($\sim 10^7$ GeV) for the MQ systems studied. On the other hand, for pions and muons, 
the decay losses dominate for lower energies, due to their short life time (see Table 2). 

The main difference between these two MQ systems (SS 433 and M33 X-7) concerns the synchrotron cooling rates. In general, 
according to Eqs. (\ref{Kin-Lumin}) and (\ref{MagnField}), for systems with wide half-opening angle (e.g. M33 X-7, with 
$\xi = 7^{\circ}$), the magnetic energy density is lower than that of systems with small $\xi$ (e.g. the SS 433 with 
$\xi=0.6^{\circ}$). Hence, the magnetic field for M33 X-7 is lower which becomes obvious by comparing Eq. (\ref{MagnField}) 
for the two systems. The latter conclusion justifies the lower synchrotron loss rate for the M33 X-7 system.

\vspace*{0.20 cm}

\section{ Formulation of the key-role transport equations inside jet plasmas }
\label{Model}

In this section, we make a first step towards improving the mathematical formulation describing the creation, scattering, 
decay, diffusion and emission of particles (including neutrinos) and electromagnetic radiation in astrophysical outflows 
(jets). The basic key-role tool of this formalism is the general transport equation which satisfies each of the particles 
considered \citep{Vieyro-Romero}. 

We should note that, such a formalism may be reliably applicable for multi-species (multi-particle) emissions from MQs and 
X-ray binaries, as well as from central regions of galaxies like the AGN systems, in which the compact object is a massive 
or supermassive black hole, see e.g. \citep{Romero_Review,Ginzburg}. The geometry of the latter sources is assumed to be 
rather spherical, while in the first class of systems that are studied in this work, the sources are assumed conical and 
the particles (neutrinos) or radiation are emitted to the forward or backward direction. 

From a mathematical point of view, the transport equation which satisfies any kind of particles moving into the (dark) 
jets of MQs, is an integro-differential equation, as is explained below.

\subsection{The transport equation for particles moving inside MQ jets} 

The general transport equation describes the concentration (distribution) of particle of j-kind, $N_j(E,{\bf r},t)$, 
where $j = p, e^\pm, \pi^\pm, \mu^\pm$, etc., as a function of the time $t$, the particle's energy $E$, and the position 
${\bf r}$ inside the (conical) jet. In essence, this is a phenomenological macroscopic equation of particle (or radiation) 
transport describing astrophysical outflows, and it is written as

\begin{strip}
\begin{equation}
\frac{\partial N_j}{\partial t} - \nabla\cdot\left( D_j\nabla N_j\right) + 
\frac{\partial\left(b_j N_j\right)}{\partial E}-\frac{1}{2}\frac{\partial^2\left(d_j N_j\right)}{\partial E^2}
= Q_j(E,{\bf r},t) - p_jN_j + \sum_k \int P_j^k (E^\prime,E) N_j(E,{\bf r},t) dE \, , \quad j = p, \pi^\pm, \mu^\pm \, ...
\label{Gen_Trans_Equ}
\end{equation}
\end{strip}
\noindent
where the parameters $D_j$, $b_j$, $d_j$, $p_j$ and $P_j^k$ may depend upon the space and time coordinates and also on the
energy $E$. The latter equation coincides with the continuity equation for particles of type-$j$, with j=1,2,3 and
$(1,2,3)\equiv$(p, $\pi$, $\mu$)]. 

The term $Q_j(E,{\bf r},t)$ in the r.h.s. of Eq. (\ref{Gen_Trans_Equ}) is equal to the intensity of the source producing the 
particles-$j$, which is also known as the injection function of particles-j. This means that $Q_j(E,{\bf r},t)dEd^3{\bf r}dt$ 
represents the number of particles kind-$j$ provided by the sources in a volume element $d^3{\bf r}$, in the energy range between 
$E$ and $E+dE$ during the time $dt$. In the case when the $j$-type particles are products of a chain reaction (as it holds in 
our present work assuming the p-p reaction chain described in Sect. 2.2), the function $Q_j(E,{\bf r},t)$ couples the $j$-reaction 
with its parent reaction, i.e. the equation of particles kind-(j-1).

Further, the term proportional to $p_j$ exists in cases when catastrophic (or knock out) processes take place that cause 
catastrophic energy losses. Then, this gives the probability per unit time for the losses in question to occur. Thus, the 
coefficient $p_j$ is written as
\begin{equation}
p_j = \frac{1}{T_j} \equiv t^{-1}_j \, ,
\end{equation}
with $T_j$ is the mean life time of the particles of kind-$j$. Also $p_jN_j$ denotes the number of particles "knocked out" 
per unit time. 

The coefficients $b_j$ in the third term (r.h.s.) are equal to the mean energy increment of the particle-$j$ per unit time i.e.
\begin{equation}
b_j = \frac{dE}{dt} \, .
\end{equation}
For the special case when $b_j = b_j(E)$, this coefficient is related to energy losses of various cooling processes. Also, the 
coefficients $P^k_j(E)$, with $k > j$, in the latter summation of the transport equation, are related with the existence of 
fragmentation of the primary (or secondary) particles which in this work are assumed as non-existing.

The parameter $D_j$ is known as the diffusion coefficient which, in general, is a functions of the coordinates $\bf r$ 
and time $t$ if the concentration of the jet plasma and its macroscopic motion are inhomogeneous in the volume of the jet.
By assuming that $D_j = 0$ we may reliably describe the regular motion of particles j-kind along the lines of force of 
the magnetic field ${\bf B}$.

For simplicity, in this work we make the realistic assumptions that, the above coefficients depend only on the particle's
energy $E$, i.e. we assume that $D_j(E)$, $b_j(E)$ and $p_j(E)$. Then, the system of general transport equations (\ref{Gen_Trans_Equ}) 
reads

\begin{strip}
\begin{equation}
\frac{\partial}{\partial t} 
\begin{pmatrix} 
N_p(E,z,t) \\ N_{\pi}(E,z,t) \\ N_{\mu}(E,z,t)
\end{pmatrix}
- \nabla^2 \begin{pmatrix}
D_p(E) N_p(E,z,t) \\ D_{\pi}(E) N_{\pi}(E,z,t) \\ D_{\mu}(E) N_{\mu}(E,z,t) 
\end{pmatrix}
+ \frac{\partial}{\partial E}
\begin{pmatrix} 
b_p(E) N_p(E,z,t) \\ b_{\pi}(E) N_{\pi}(E,z,t) \\ b_{\mu}(E) N_{\mu}(E,z,t) 
\end{pmatrix}
-\frac{1}{2}\frac{\partial^2}{\partial E^2}
\begin{pmatrix} 
d_p(E) N_p(E,z,t) \\ d_{\pi}(E) N_{\pi}(E,z,t) \\ d_{\mu}(E) N_{\mu}(E,z,t) 
\end{pmatrix}
+ \begin{pmatrix} 
t_{esc}^{-1} N_p(E,z,t) \\ t_{\pi}^{-1} N_{\pi}(E,z,t) \\ t_{\mu}^{-1} N_{\mu}(E,z,t) 
\end{pmatrix}
= \begin{pmatrix} 
Q_p(E,z,t) \\ Q_{\pi}(E,z,t) \\ Q_{\mu}(E,z,t) 
\end{pmatrix}
\label{Syst_transp_Equat}
\end{equation}
\end{strip}
Note that, in the latter system of coupled differential equations we write down only three particles (protons, 
pions and muons), without considering the complete reaction family tree, since we haven't distinguished particles 
with different charge as $\pi^\pm$ and $\mu^\pm$, and we haven't considered the possibility of left-right symmetry, 
$\mu_{L,R}$ of muons as we have done below in Sect. 4. This means that, by considering all these particles, the 
system of Eq. (\ref{Syst_transp_Equat}) will have seven lines.
 
Usually, we ignore the term involving second derivative with respect to $E$ means that the acceleration term is negligible.

The solution of the general transport equation for particles of one kind (when the last term of this equation can 
be omitted) is described in detail in Ref. \citep{Ody-Smpon-2018}. In the latter work, however, neutrinos produced 
only from $\pi^\pm$ have been considered. In the present paper we proceed further and include neutrino emissions also 
from the muon decays, so we need to solve the system of Eq. (\ref{Syst_transp_Equat}) by including the third line too. 

The complete form of the system of Eq. (\ref{Syst_transp_Equat}) is treated with the method of Ref. \citep{Tsoulos_Ody_Stavrou} 
in order to find the exact solutions. In Ref. \citep{PapOKos}, equation (\ref{Syst_transp_Equat}) is treated 
semi-analytically.
In the present work, we restrict ourselves to simplified forms resulting by neglecting various phenomena (processes) 
taking place inside the astrophysical outflows (jet plasma), even though some of the assumed omissions may be considered 
as rather crud approximations. 

Below we discuss the cases of Eq. (\ref{Syst_transp_Equat}) satisfying the conditions of the one-zone approximation 
\citep{Khangulyan,Ody-Smpon-2018}, i.e. the cases when the particle distributions are independent of time (steady state 
approximation).

\subsubsection{ Transport equation assuming absence of energy-losses } 

At first, in the calculations of our present work, we start by writing the simplest solutions of the system of transport 
equations obtained by assuming that all energy losses are absent i.e. $b_j = 0$. Then, by denoting $N_{p,0}$, $N_{\pi,0}$ 
and $N_{\mu,0}$ the corresponding energy distributions for protons, pions and muons, respectively, the system of transport 
equations takes the trivial form
\begin{equation}
\begin{pmatrix} 
t_p^{-1}(E) N_{p,0}(E,z) \\ t_\pi^{-1}(E) N_{\pi,0}(E,z) \\ t_\mu^{-1}(E) N_{\mu,0}(E,z)
\end{pmatrix} 
=
\begin{pmatrix} Q_p(E,z) \\ Q_\pi(E,z) \\ Q_\mu(E,z) \end{pmatrix}
\end{equation}
The calculated distributions $N_{j,0}$, with j=p, $\pi^\pm$, $\mu^\pm$ of the latter equations are discussed in the next 
Section.

\subsubsection{Steady state transport equations with particle losses when moving inside the jet plasma}

Under the conditions of the steady state approximation, and assuming that the various energy losses are absent, the system 
of transport equations Eq. (\ref{Syst_transp_Equat}) takes the form
\begin{equation}
\frac{\partial}{\partial E}
\begin{pmatrix} 
b_p(E) N_p(E,z) \\ b_{\pi}(E) N_{\pi}(E,z) \\ b_{\mu}(E) N_{\mu}(E,z) 
\end{pmatrix}
+ \begin{pmatrix} 
t_{esc}^{-1} N_p(E,z) \\ t_{\pi}^{-1} N_{\pi}(E,z) \\ t_{\mu}^{-1} N_{\mu}(E,z) 
\end{pmatrix}
= \begin{pmatrix} 
Q_p(E,z) \\ Q_{\pi}(E,z) \\ Q_{\mu}(E,z) 
\end{pmatrix}
\label{Steady_transp_Eqs}
\end{equation}
%
In order to calculate the time independent neutrino and gamma-ray emissivities, we need, first, to calculate the 
distributions of protons, pions and muons, $N_p$, $N_{\pi}$, $N_{\mu}$, respectively, from Eq. (\ref{Steady_transp_Eqs}). 
For these computations we used a code written in the C programming language, mainly following the assumptions of Refs. 
\citep{Reynoso08,Reynoso09,Vieyro-Romero}. 

In the latter case, the second term of the l.h.s. of the transport equation we replace the escape rate, $t_{esc}^{-1}$ 
of Eq. (\ref{t_esc}) and $b(E)= - E t_{loss}^{-1} (E) \,$. The latter quantity the energy loss rate for each particle 
which is given by
\begin{eqnarray}
& b_p(E)    & = - E t_{loss,p}^{-1} (E) = -E(t_{syn}^{-1} + t_{ad}^{-1} + t_{pp}^{-1} ),\\
& b_{\pi}(E)& = - E t_{loss,\pi}^{-1} (E) = -E(t_{syn}^{-1} + t_{ad}^{-1} + t_{\pi p}^{-1} ),\\
& b_{\mu}(E)& = - E t_{loss,\mu}^{-1} (E) = -E(t_{syn}^{-1} + t_{ad}^{-1})
\end{eqnarray}
It is worth noting that, in the latter equations some additional terms may appear in cases when the rates of some other 
processes, ignored in our present work (like e.g. the Inverse Compton Scattering, the pion-muon scattering, etc.), may be 
considered important for the description of other cosmic structures.

\section{Injection functions and energy distributions of particles inside the jet}

In this section, the source functions $Q_j(E,{\bf r},t)$ entering the r.h.s of the system of coupled integro-differential
equations of transport type, are discussed. Towards this aim, initially in the system of Eqs. (\ref{Syst_transp_Equat}), 
we insert phenomenological expressions obtained from Ref. \citep{Lipari07}. A mathematical semi-analytic way of solving 
this system is proposed in Ref. \citep{Ody_in_progress}. Also, following Ref. \citep{Tsoulos_Ody_Stavrou}, a method is 
derived for the numerical solution of the differential system of transport equations. 

\subsection{Proton injection function and energy distribution }

\subsubsection{Relativistic proton injection function }

In the assumed p-p mechanism, the density of fast (relativistic) protons injected is more important, while the 
corresponding density of slow (thermal) protons is dynamically significant.

A usual injection function for the relativistic protons, coming out of the acceleration mechanism, has been taken to 
be a power-law with exponent equal to two, i.e. a function of the energy of the form \citep{Ghisellini}
\begin{eqnarray}
Q_{p}(E,z) = Q_0 \left(\frac{z_0}{z}\right)^3 \frac{1}{E^2} \, ,
\label{Prot_Rel-Inj-Fun_powerlaw}
\end{eqnarray}
where $Q_0$ is a normalization constant obtained by specifying the power in the relativistic protons \citep{Reynoso08}, 
see Appendix. The above form is valid for the jet's frame of reference (the corresponding expression in observer's 
system is shown in Appendix). 

It should be noted that, Eq. (\ref{Prot_Rel-Inj-Fun_powerlaw}) enters (through integration) the determination of the 
injection functions of the secondary particles ($\pi^\pm$, $\mu^\pm$, see below) and this justifies the term systems 
of coupled integro-differential equations used in Sect. 3. The latter injection functions are also dependent on the 
rates and cross sections of the reactions that preceded in the chain processes following the inelastic p-p collision.

\subsubsection{Relativistic Proton energy distribution in the jet}

Under the assumptions discussed before, the system of transport equations (\ref{Steady_transp_Eqs}) can be easily solved 
and the obtained proton distribution $N_p(E,z)$ is written as \citep{Reynoso08,Romero-Vila-08}
\begin{equation}
N_p(E,z) = \int_{E_p}^{E_p^{max}} \mid b_p(E) \mid^{-1} Q_p(E',z)e^{-\tau_p (E,E')}dE'
\label{Prot_Ener_Distr}
\end{equation}
where
\begin{equation}
\tau_p (E,E') = \int_{E}^{E'} \frac{t_{esc,p}^{-1} d E^{\prime \prime}}{\mid b_p(E^{\prime \prime})\mid} \, .
\label{tau_proton}
\end{equation}
The quantity $Q_p(E,z)$ corresponds to the relativistic injection function of protons at the observer's frame (see 
Appendix). The minimum energy of protons is $E_p^{min}=E_{th} = 1.22$ GeV, while the maximum energy is assumed to be 
$E_p^{max}= 10^6 - 10^7$ GeV.

\begin{figure}
\includegraphics[width=0.6\linewidth]{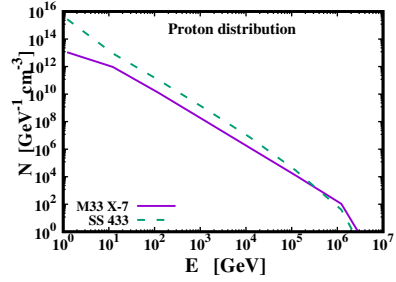}
\caption{Proton energy distribution for M33 X-7 and SS 433 microquasar systems.
\label{Pion_distr}}
\end{figure}

Figure \ref{Pion_distr} shows the proton energydistribution $N_p(E)$ at the base of the jet for M33 X-7 (solid line) and 
SS 433 (dashed line). As can be seen, the number of protons appears reduced (even by two orders of magnitude) for the 
wide half-opening angle system of M33 X-7 ($\xi = 7^{\circ}$) as compared to the narrow of SS 433 ($\xi=0.6^{\circ}$). 
This is because, the jet radius $r(z)$ is bigger and hence the magnetic field $B$ is smaller in the case of M33 X-7 than 
that of SS433.

\subsection{Pion injection functions and energy distributions inside the jet}

\subsubsection{Pion injection functions}

For pions produced through the inelastic p-p scattering, as injection function we adopt the one given by \citep{Kelner1} 
as
\begin{eqnarray}
Q_{\pi}(E,z)=n(z)c\int_{\varepsilon}^{1} \sigma_{pp}^{inel} \left(\frac{E}{x}\right) N_p
\left(\frac{E}{x},z\right) 
F_{\pi}\left(x,\frac{E}{x}\right) \frac{dx}{x} 
\end{eqnarray}
where $\varepsilon = E/E_p^{max}$, $x=E/E_p$ and $F_\pi$ denotes the distribution of pions produced per $p-p$ collision 
(see Appendix).

\subsubsection{Pion energy distributions}

The steady state energy distribution $N_{\pi}(E,z)$ for pions (created through the scattering of hot protons off the cold 
ones), results through a similar way to that of Eqs. (\ref{Prot_Ener_Distr}) and (\ref{tau_proton}). With the replacement 
$p\rightarrow \pi$ and $t_{esc}^{-1}\rightarrow t_{\pi}^{-1}(E,z)$, respectively, on the latter equations the corresponding 
solution for pion energy distribution is, then, written as
\begin{eqnarray}
N_{\pi}(E,z) & = \int_{E}^{E_{max}} \mid b_{\pi}(E)\mid^{-1} Q_{\pi}(E',z)e^{-\tau_{\pi}}dE' 
\label{pion_Ener_Distr}
\end{eqnarray}
where the rate $t_{\pi}^{-1}$ includes contributions from the decay and escape rates [see Eq. (\ref{pion-muon-rate})].

\subsection{Muon injection functions and energy distributions inside the jet}

\subsubsection{Muon injection functions}

As mentioned before, in this work in addition to neutrinos coming from the decay of charged pions $\pi^\pm$, we evaluate 
neutrino emissivities originating from the decay of the secondary charged muons $\mu^\pm$. For the latter 
emissivity, we follow Ref. \citep{Lipari07}, in order to take into account properly the muon energy loss. This means that,
it is necessary to consider the production of both left handed and right handed muons, $\mu_L^-$ and $\mu_R^+$, separately, 
because $\mu_L^-$ and $\mu_R^+$ have different decay spectra. 

Thus, the injection functions of the left handed and right handed muons are \citep{Lipari07}:
\begin{eqnarray}
Q_{\mu_L^-,\mu_R^+}(E_{\mu},z) = \int_{E_{\mu}}^{E^{max}} & t_{dec,\pi}^{-1} (E_{\pi}) N_{\pi}(E_{\pi},z)
\Theta(x-r_{\pi})  \nonumber\\ 
& \times \frac{r_{\pi}(1-x)}{E_{\pi}x{(1-r_{\pi})}^2} d E_{\pi}
\end{eqnarray}
and
\begin{eqnarray}
Q_{\mu_R^-,\mu_L^+}(E_{\mu},z)  = \int_{E_{\mu}}^{E^{max}} & t_{dec,\pi}^{-1} (E_{\pi}) N_{\pi}(E_{\pi},z)
\Theta(x-r_{\pi}) \nonumber\\
& \times \frac{(x-r_{\pi})}{E_{\pi}x{(1-r_{\pi})}^2}  d E_{\pi}
\end{eqnarray}
with $x = E_{\mu}/E_{\pi}$ and $r_{\pi} = {(m_{\mu}/m_{\pi})}^2$.

\subsubsection{Muon energy distributions}

The corresponding muon energy distributions, coming out of the transport equation, result from Eq. (\ref{pion_Ener_Distr}) 
via the replacement $t_{esc}^{-1}\rightarrow t_{\mu}^{-1}(E,z)$, for muons and are then written as
\begin{eqnarray}
N_{\mu_i}(E,z) & = \int_{E}^{E_{max}} \mid b_{\mu}(E)\mid^{-1} Q_{\mu_i}(E',z)e^{-\tau_{\mu}}dE' 
\label{muon_Ener_Distr}
\end{eqnarray}
where the rate $t_{\mu}^{-1}$ for muons is given in Eq. (\ref{pion-muon-rate}).
\begin{figure}
 \includegraphics[width=0.6\linewidth]{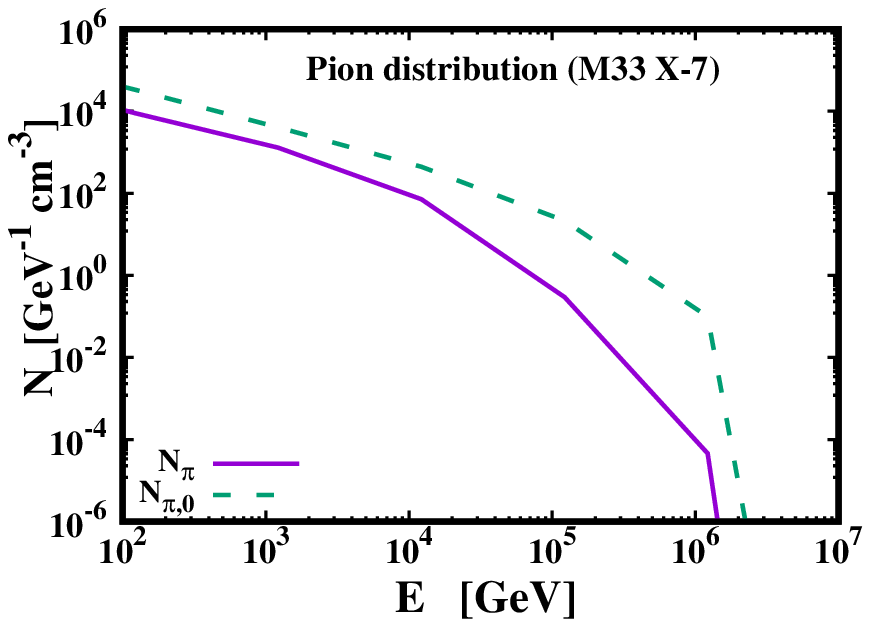} \\ 
 \includegraphics[width=0.6\linewidth]{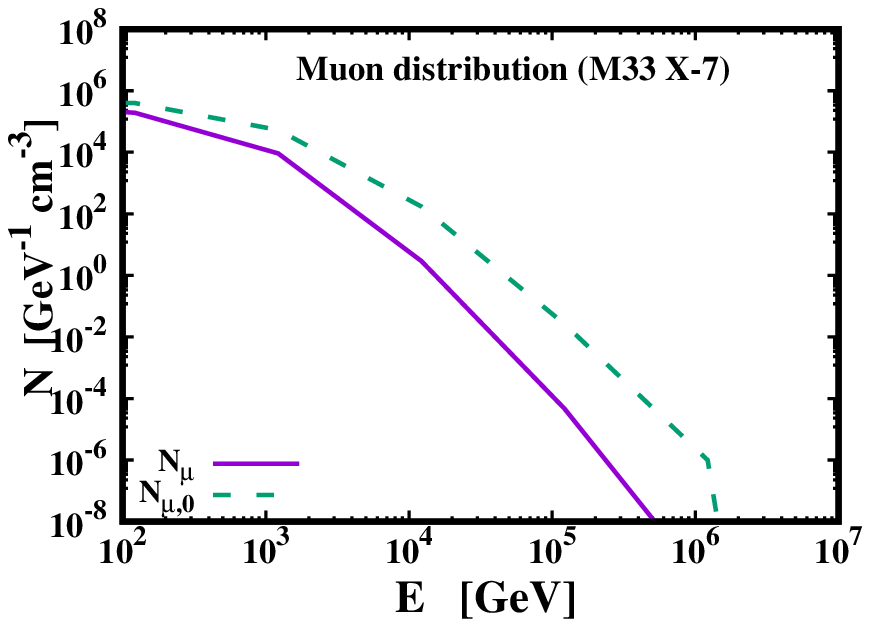}
\vspace*{0.2 cm}
\caption{Pion (Top) and muon (bottom) distributions for M33 X-7. In both case of particles-j, we compare the distribution which 
considers the important energy losses $N_{j}(E,z=z_0)$ and with that which neglects energy losses $N_{j,0}(E,z=z_0)$ for $j =\pi, \ \mu$. 
Both types of particle distributions refer to the base of the jet, i.e. at $z=z_0$ 
\label{Pi-Mu_distr}}
\end{figure}

In Fig. \ref{Pi-Mu_distr}, the pion (top) and muon (bottom) energy distributions for M33 X-7 are illustrated. In both cases 
of particles, the distribution $N_{j}(E,z=z_0)$, with $j =\pi$ or $\mu$, which considers the most important energy losses, 
is compared with that which neglects energy losses, $N_{j,0}(E,z=z_0)$. The solid lines correspond to distributions obtained 
by considering energy losses and the dashed lines correspond to those where the energy losses obtained have been neglected.

\section{Results for neutrino and $\gamma$-ray emissivities }
\label{Results neutrino and gamma ray}

In this Section, we present and discuss simulated emissivities of high energy neutrinos and $\gamma$-rays produced in the 
extragalactic microquasar M33 X-7 system which is located in the Messier 33 galaxy, at a distance $\sim 840 - 960$ kpc from 
the Earth \citep{PietschWC}. The derived algorithms for this purpose have been tested on the well-studied Galactic micro-quasars 
SS 433 and Cyg X-1 system as well.

Because in our previous calculations on neutrino production from MQs \citep{Ody-Smpon-2015,Ody-Smpon-2017,Ody-Smpon-2018}, we neglected 
(due to the complexity and long time consuming) emissivity (intensity) of neutrinos originated from the secondary muons ($\mu^{\pm}$)
produced from the charged pion ($\pi^\pm$) decays, in this section, we present contributions also originating from the $\mu^{\pm}$ 
channel [see Eq. (\ref{mu-decay})].

\begin{figure*}
\begin{center}
  \includegraphics[width=0.4\linewidth]{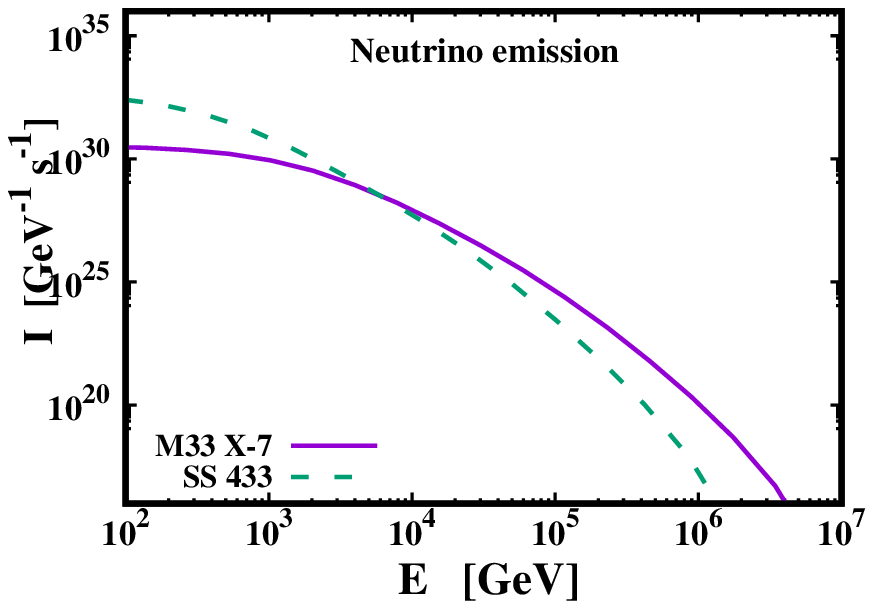}
  \hspace*{-0.5 cm}
  \includegraphics[width=0.4\linewidth]{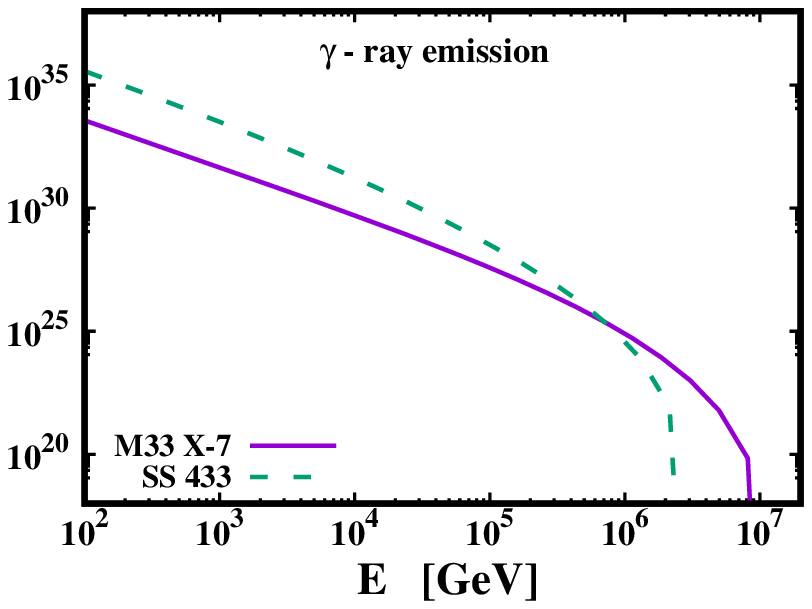}
  \caption{Neutrino (left) and gamma-ray (right) intensity for M33 X-7 and SS433
  \label{Neutr-Gamma_Emiss}}
\end{center}
\end{figure*}

By using the concentrations for protons $N_p$, pions $N_\pi$ and muons $N_\mu$, the neutrino and $\gamma$-ray intensities 
are subsequently calculated as described below.

\subsection{ Neutrino emission from the p-p reaction family tree }

After the above discussion, the total emissivity $Q_{\nu}(E,z)$ produced from a MQ is the sum of contributions from the 
two sources: 
 (i) the first comes from the direct $\pi^\pm$ decay (prompt neutrino production), and 
(ii) the second comes from the $\mu^\pm$ decay (delayed neutrino production). Thus,
\begin{equation}
Q_{\nu}(E,z)=Q_{\pi\rightarrow\nu}(E,z) + Q_{\mu\rightarrow\nu}(E,z)
\end{equation}  
For pion decays the injection function is given by
\begin{eqnarray}
Q_{\pi\rightarrow\nu}(E,z)=\int_{E}^{E_{max}} t_{dec,\pi}^{-1}(E_{\pi})N_{\pi}(E_{\pi},z)
\frac{\Theta (1-r_{\pi}-x)}{E_{\pi}(1-r_{\pi})}dE_{\pi}
\label{pi-nu_emissiv}
\end{eqnarray}
with $x=E/E_\pi$, while for the four types of muon decays the injection function reads
\begin{eqnarray}
&Q_{\mu\rightarrow\nu}(E,z) = \sum_{i=1}^{4} \int_{E}^{E_{max}} t_{dec,\mu}^{-1}(E_{\mu}) N_{\mu_i}(E_{\mu},z)\nonumber\\
& \times \left[\frac{5}{3}-3x^2+\frac{4}{3}x^3+\left(3x^2-\frac{1}{3}-\frac{8}{3}x^3\right)h_i\right]\frac{dE_{\mu}}{E_{\mu}} \, .
\label{mu-nu_emissiv}
\end{eqnarray}
(with $x=E/E_\mu$). 

In the last expression, the symbols $\mu_i, i=1,2,3,4$ correspond to $\mu_{\lbrace 1,2\rbrace} = \mu_L^{\lbrace -,+\rbrace}$, 
$\mu_{\lbrace 3,4\rbrace} = \mu_R^{\lbrace -,+\rbrace}$ \citep{Lipari07}, while $h_{\lbrace1,2\rbrace}=-h_{\lbrace3,4\rbrace}=-1$. 
Then, the calculation of each of the latter two integrals of Eqs. (\ref{pi-nu_emissiv}) and (\ref{mu-nu_emissiv}) provides 
separately the partial emissivity of neutrinos for the prompted and the delayed neutrino source, respectively. Needless to
note that Earth and space telescope are not able to discriminate the two source as it happens with laboratory neutrino sources.

Subsequently, we easily obtain the neutrino intensity (in units $GeV^{-1} s^{-1}$) by the spatial integration
\begin{equation}
I_{\nu}(E)=\int_V Q_{\nu}(E,z)d^3r =\pi \left( \tan \xi\right)^2 \int_{z_0}^{z_{max}} Q_{\nu}(E,z)z^2dz \, .
\end{equation}

Figure \ref{Neutr-Gamma_Emiss} (left) shows the neutrino intensity produced at the base of the jet from direct decays of 
secondary pions and muons coming from p-p collisions in the jets of M33 X-7 and SS 433, respectively. As can be seen, the 
number of produced neutrinos reduces significantly for energies $E>10^3$ GeV, following the behavior of pion and muon
distribution. We can also see that the neutrinos produced in M33 X-7 are less than those produced in SS 433, due to the 
wider-half opening angle $\xi$ as we explained before. 

It should be noted that, to perform the above calculations we have updated and improved our codes to reduce the time 
consuming to a reasonable level, so as the integral involved in Eq. (\ref{mu-nu_emissiv}) to be easily obtained, since 
in going the step from Eq. (\ref{pi-nu_emissiv}) to Eq. (\ref{mu-nu_emissiv}) the time consuming increases rapidly.

\subsection{ Gamma-ray emission from the p-p reaction chain }

As we have discussed in Sect. 2, the p-p collisions inside the jets, produce secondary neutral particles (pions, eta particle, 
etc.) that decay to give $\gamma$-rays. The dominant of these channels goes through the reaction 
\begin{equation}
\pi^0 \rightarrow \gamma + \gamma \, .
\end{equation}
For $E_{\gamma}\geq 100$ $GeV$, we consider the $\gamma$-ray emissivity at a distance $z$ along the jets (in units 
$GeV^{-1} s^{-1}$) to be given by \citep{Reynoso08} as
\begin{equation}
Q_{\gamma}= c \int_{E_\gamma/E_p^{max}}^1\sigma_{pp}^{inel}\left(\frac{E_\gamma}{x}\right) 
N_p\left(\frac{E_{\gamma}}{x},z\right)F_{\gamma}\left(x,\frac{E_\gamma}{x}\right)\frac{dx}{x} \, .
\end{equation}
where $F_{\gamma}$ is the spectrum of the produced $\gamma$-rays \citep{Kelner1, Kelner2} with energy $x=E_{\gamma}/E_p$ 
for a primary proton energy $E_p$. The function $F_{\gamma}$ is given in Appendix.

Subsequently, the corresponding spectral intensity of $\gamma$-rays, can be obtained from the following spatial integration 
over the jet's volume $V$ as
\begin{equation}
I_{\gamma}(E_{\gamma})= \int_V Q_{\gamma} (E_\gamma,z)d^3r = \pi \left(\tan \xi\right)^2 \int_{z_0}^{z_{max}} Q_{\gamma} 
(E_\gamma,z)z^2dz,
\end{equation}

Figure \ref{Neutr-Gamma_Emiss} (right) shows the $\gamma$-ray intensity for energies $E_{\gamma} > 100$ GeV. As can be 
seen, the produced $\gamma-ray$ intensity reduces rather steadily following the behavior of the proton distribution in 
both cases of MQ systems. Again for the M33 X-7 it is lower mainly due to the wider half-opening angle (i.e. weaker 
magnetic field) but also due to other parameters. 

In our ongoing calculations \citep{Ody_in_progress}, among others we include results obtained by using the 3-D PLUTO 
astrophysical hydrocode where the jets of the studied systems are approximated as purely relativistic magnetohydrodynamical 
(RMHD) flows with the magnetic field (assumed tangled) being dragged with the flow \citep{Smponias_tsk_2014}.

\section{Summary and Conclusion}

In this work we address neutrino and $\gamma$-ray emissions from microquasars and X-ray binary stars (XRBs) that consist 
of a stellar mass black hole (compact object) and a main sequence donor star. We assume that these emissions originate 
from decay and scattering processes of the secondary particles produced through the $p-p$ scattering mechanism, i.e. the 
inelastic collision of relativistic protons of the jet with the thermal ones.

Such high energy neutrinos and $\gamma$-rays are detectable by operating terrestrial and space telescopes. Among the 
operating detectors are the under ice IceCube (at the South Pole), the ANTARES and KM3Net (under the Mediterranean sea), 
and many others.

In the near future, more accurate measurements of gamma-ray and neutrino fluxes by the coming observatories, such as 
the CTA (Acharya et al.2018), which is expected to shed more light on the nature and the emission sources. From the 
perspective of neutrino detection, the addition of more years of data with continuous operation of IceCube will improve 
the sensitivity of the search for Galactic sources of cosmic neutrinos. Furthermore, the next generation detection
instrument, IceCube-Gen2, a substantial expansion of IceCube, will be 10 times larger. This next-generation neutrino 
observatory with five times the effective area of IceCube is expected to improve the neutrino source search sensitivity 
by the same order \citep{IceCube-Gen2}. With higher neutrino statistics, identifying Galactic sources will become more 
promising.

On the other hand, theoretically, by modelling the solution of the system of coupled transport equations, we were able to
perform detailed calculations for various processes taking place inside the jets of Galactic and extra galactic system 
M33 X-7 assuming hadronic content in their jets. These cooling rates enter the proton, pion and muon energy distributions 
through which one obtains neutrino and $\gamma$-ray intensities. 

Our near future calculations include, among others, multidimentional simulations using the PLUTO astrophysical hydrocode 
which may provide both radiative and dynamical description of the astrophysical outflows within multi-scale and 
multi-messesger astrophysics.

\section*{Acknowledgements}

This research is co-financed (O.T.K.) by Greece and the European Union (European Social Fund-ESF) through the Operational Programme 
``Human Resources Development, Education and Lifelong Learning 2014- 2020" in the context of the project (MIS5047635). D.A.P. wishes 
to thank Prof. T.S. Kosmas for fruitful discussions during his stay in the Dept. of Physics, University of Ioannina.


\section*{Data Availability}

There is no data used in this paper.





\appendix

\section{ }

\subsection{\bf The hot proton's injection function}
\label{appendix:a}
At the observer's frame of reference, the injection function of protons, $Q_p(E,z)$, and is given by \citep{Reynoso08}:
\begin{eqnarray}
Q_p(E,z)= \left(\frac{z_0}{z}\right)^3 \frac{Q_0}{\Gamma_b \left(E-\beta_b \cos\theta\sqrt{E^2-m^2c^4} \right)^2} \nonumber \\
 \times \left[1-\frac{\beta_bEcos\theta}{\sqrt{E^2-m^2c^4}}\right]
\end{eqnarray}
where $\Gamma_b$ is the bulk Lorentz factor of the jet, and $\theta$ denotes the angle between the jet's injection axis
and the direction of the line of sight (LOS).

Then, the normalization constant $Q_0$ is obtained by determining the power $L_p$ of the relativistic protons \citep{Reynoso08},
given by
\begin{equation}
L_p=\int_{V}d^3r\int_{E_p^{min}}^{E_p^{max}}E_pQ_p(E_p,z)dE_p
\end{equation}

After performing the latter integration, the normalization constant $Q_0$ (see text) reads \citep{Reynoso08}
\begin{equation} 
Q_0 = \frac{2 c}{z_0} K_0 , \quad  \quad K_0=\frac{4q_{rel} L_k}{cr_0^2ln\left({E_p'^{max}/E_p'^{min}}\right)} \, .
\end{equation}

\subsection{\bf The distribution of pions produced per p-p collision}
\label{appendix:b}

The distribution of pions produced per p-p collision is
\begin{eqnarray}
&F_{\pi}^{(pp)}\left(x.\frac{E}{x}\right)=4\alpha B_{\pi}x^{\alpha -1}\left(\frac{1-x^{\alpha}}{1+rx^{\alpha}(1-x^{\alpha})}\right)^4\nonumber\\
& \times \left(\frac{1}{1-x^{\alpha}}+\frac{r(1-2x^{\alpha})}{1+rx^{\alpha}(1-x^{\alpha})}\right)\left(1-\frac{m_{\pi}c^2}{xE_p}\right)^{\frac{1}{2}}
\end{eqnarray}
with
$x=E/E_p$, $B_\pi=\alpha '+ 0.25$, $\alpha '=3.67+0.83L+0.075L^2$, $r= 2.6/\sqrt{\alpha '}$, and $\alpha=0.98/\sqrt{\alpha '}$ 
\citep{Kelner1,Kelner2}.

\subsection{\bf Produced $\gamma$-ray spectrum}
\label{appendix:c}

Following the treatment of \citep{Kelner1, Kelner2}, the spectrum of produced gamma-rays with energy $x={E_\gamma}/{E_p}$ for a primary proton with energy $E_p$ is written as
\begin{eqnarray}
F_{\gamma}(x,E_p)&=&B_{\gamma}\frac{lnx}{x}\left[\frac{1-x^{\beta_{\gamma}}}{1+k_{\gamma}x^{\beta_{\gamma}}(1-x^{\beta_{\gamma}})}\right]^4\\
& \times & \left[\frac{1}{lnx}-\frac{4\beta_{\gamma}x^{\beta_{\gamma}}}{1-x^{\beta_{\gamma}}}-\frac{4k_{\gamma}\beta_{\gamma}x^{\beta_{\gamma}}(1-2x^{\beta_{\gamma}})}{1+k_{\gamma}x^{\beta_{\gamma}}(1-x^{\beta_{\gamma}})}\right] \nonumber
\end{eqnarray}
where
\begin{eqnarray}
&B_{\gamma}&=1.3+0.14L+0.011L^2,\nonumber\\
&\beta_{\gamma}&=\frac{1}{0.008L^2+0.11L+1.79},\\
&k_{\gamma}&=\frac{1}{0.014L^2+0.049L+0.801}\nonumber
\end{eqnarray}
with
\begin{equation}
L=ln\left( E_p/ (1 TeV) \right)
\end{equation}

\bsp	
\label{lastpage}
\end{document}